\documentclass[12pt]{article}
\usepackage{amsmath}
\usepackage{txfonts}

\newcommand{\cak}{c_{1 \mathbf k}}
\newcommand{\cbk}{c_{2 \hskip 0.2 mm \mathbf k}}
\newcommand{\cjk}{c_{j \hskip 0.2 mm \mathbf k}}
\newcommand{\cakd}{c^{\dagger}_{1 \mathbf k}}
\newcommand{\cbkd}{c^{\dagger}_{2 \mathbf k}}
\newcommand{\cjkd}{c^{\dagger}_{j \hskip 0.2 mm \mathbf k}}
\newcommand{\clkd}{c^{\dagger}_{l \mathbf k}}
\newcommand{\coslk}{\cos \left( \cfrac{\Lambda_{\mathbf k} t}{\hbar} \right)}
\newcommand{\ccoslk}{\cos^2 \left( \cfrac{\Lambda_{\mathbf k} t}{\hbar} \right)}
\newcommand{\cosllk}{\cos \left( \cfrac{2 \Lambda_{\mathbf k} t}{\hbar} \right)}
\renewcommand{\d}{\mathrm d}
\newcommand{\e}{\mathrm e}
\newcommand{\fbraket}[1]{
   \mathinner{\langle{#1}\rangle} 
}
\newcommand{\Fj}{F_{\! j}}
\newcommand{\half}{\frac{1}{2}}
\newcommand{\Htot}{\hat{\mathrm H}}
\newcommand{\Hpert}{\hat{\mathrm H}^{\prime}}
\newcommand{\Hunpert}{\hat{\mathrm H}_0}
\newcommand{\Hunperta}{\hat{\mathrm H}_{01}}
\newcommand{\Hunpertb}{\hat{\mathrm H}_{02}}
\newcommand{\ID}{I_{\mathrm D}}
\newcommand{\im}{\mathrm i}
\newcommand{\JD}{J_{\mathrm D}}
\newcommand{\kB}{k_{\mathrm B}}
\newcommand{\Nop}{\hat{\mathrm N}}
\newcommand{\rhoh}{\hat{\rhoup}}
\newcommand{\sinlk}{\sin \left( \cfrac{\Lambda_{\mathbf k} t}{\hbar} \right)}
\newcommand{\Sp}{                 
   \mathbf{\mathsf{Tr}}
}
\newcommand{\sinllk}{\sin \left( \cfrac{2 \Lambda_{\mathbf k} t}{\hbar} \right)}
\newcommand{\ssinlk}{\sin^2 \left( \cfrac{\Lambda_{\mathbf k} t}{\hbar} \right)}
\newcommand{\epsk}{\epsilon_{\mathrm k}}
\newcommand{\lk}{\Lambda_{\mathrm k}}
\newcommand{\lkv}{\Lambda_{\mathbf k}}
\newcommand{\sumk}{\sum_{\mathbf k}}
\newcommand{\veck}{\mathbf k}
\title{Quantum diffusion: a simple, exactly solvable model}
\author{Wim Magnus\footnote{wim.magnus@ua.ac.be}, \\
        \small {\itshape Imec, Kapeldreef 75, B-3001 Leuven, Belgium and} \\
        \small {\itshape Universiteit Antwerpen, Physics Department,} \\
        \small {\itshape Groenenborgerlaan 171, B-2020 Antwerpen, Belgium} \\
        and \\
        Kwinten Nelissen, \\
        \small {\itshape Universiteit Antwerpen, Physics Department,} \\
        \small {\itshape Groenenborgerlaan 171, B-2020 Antwerpen, Belgium}
        }
\date{ \ }
\begin{document}
\maketitle
\begin{abstract}
We propose a simple quantum mechanical model describing the time dependent
diffusion current between two fermion reservoirs that were initially
disconnected and characterized by different densities or chemical potentials.
The exact, analytical solution of the model yields the transient behavior of
the coupled fermion systems evolving to a final steady state, whereas the
long-time behavior is determined by a power law rather than by exponential
decay. Similar results are obtained for the entropy production which is
proportional to the diffusion current.
\end{abstract}
\section{Introduction}
\label{sec:intro}
Key ingredients to the investigation of non-equilibrium features of physical
systems, time irreversibility, energy dissipation and entropy production
\cite{ojima1987,ichiyanagi1989-1,ichiyanagi1989-2,ichiyanagi1989-3,kawai2007}
can be treated analytically only in a few cases. Providing a well-known example
of the latter, the Caldeira-Leggett model \cite{caldeira1983} along with all
its extensions \cite{magnus1993,wan1996,sels2012}
essentially leans on the quadratic interaction imposed to couple a huge number
of harmonic oscillators. As such, the resulting equations of motion governing
the dynamics of any relevant observables remain classical, in spite of the
quantum mechanical nature of the density matrix, potentially representing the
initial equilibrium conditions. In this light, it is tempting to conceive an
exactly solvable model that reveals a signature of quantum dynamics while
exhibiting at least one of the typical non-equilibrium features such as
irreversibility. The model presented below describes quantum diffusion as the
basic mechanism underlying the particle exchange between two fermion systems
that are brought into contact at some time instant $t = 0$, while having been
separated at all previous times. Section \ref{sec:model} introduces the model
in terms of its Hamiltonian and discusses the time evolution in the framework of
the Heisenberg picture. In section \ref{sec:soldis} the Heisenberg equations of
motion for the fermion creation and annihilation operators are explicitly solved
and the time dependent obervables, including the diffusion current, are
extracted by averaging the corresponding
Heisenberg operators with the initial density matrix characterizing the initial
equilibrium of the uncoupled fermion reservoirs. As a result, the time
dependent fermion densities are calculated explicitly and the results are
discussed in terms of the occurrence of time irreversibility. In
particular, a direct relation between the diffusion current and the entropy
production is exploited to study the time dependence of the latter. Finally, a
conclusion is drawn in section \ref{sec:concl}.

\section{Two coupled fermion systems: Heisenberg equations of motion}
\label{sec:model}
We consider two non-interacting fermion gases establishing two-dimensional
reservoirs in thermal equilibrium that are completely disconnected at all
times preceding $t = 0$, while being confined to two rectangular areas with
sizes $L_x$ and $L_y$. The uncoupled reservoirs are at the same temperature,
however with different initial densities $n_{10}$ and $n_{20}$. \\
The Hamiltonian of the disconnected 2DEGs can be written as
\begin{equation}
   \Hunpert = \Hunperta + \Hunpertb =
              \sumk \epsk \left( \cakd \cak + \cbkd \cbk \right),
\end{equation}
where $\cak, \cakd$ and $\cbk, \cbkd$ respectively denote the fermion
creation and annihilation operators for both subsystems and $\epsk$ are the
single-particle eigenergies. For the sake of simplicity, spin degrees of freedom
have been ignored whereas the eigenenergies are taken to be parabolic, i.e.
$\epsk = \hbar^2 k^2 / 2 m$ with $m$ denoting the fermion effective mass.
As of $t = 0$, the two reservoirs get connected by turning on the perturbation
\begin{equation}
   \Hpert = \sumk \lkv \left( \cakd \cbk + \cbkd \cak \right). \label{eq:Hpert}
\end{equation}
which initiates the exchange of fermions.
$\lkv$ is the transition rate specifying the number of fermions with a
given wave vector $\veck$ being transferred from one reservoir to the other.
Throughout this paper $\lkv$ is considered real and even in $\veck$.
The total Hamiltonian now reads
\begin{equation}
   \Htot = \Hunpert + \theta(t) \, \Hpert.
\end{equation}
The time dependence of the observables characterizing the diffusion process
will be extracted from the time dependent Heisenberg operators $\cak(t)$,
$\cakd(t)$, $\cbk(t)$, $\cbkd(t)$. The latter are known to obey the Heisenberg
equations of motion 
\begin{equation}
   \im \hbar \, \frac{\d \cjk(t)}{\d t} = [\cjk(t), \Htot], \quad
   \im \hbar \, \frac{\d \cjkd(t)}{\d t} = [\cjkd(t), \Htot], \quad j = 1, 2.
\end{equation}
which, in the present case, reduce to
\begin{align}
   \im \hbar \, \frac{\d \cak(t)}{\d t}
   & = \epsk \, \cak(t) + \lkv  \, \cbk(t), \\
   \im \hbar \, \frac{\d \cbk(t)}{\d t}
   & = \lkv  \, \cak(t) + \epsk \, \cbk(t),
\end{align}
because the perturbation (Eq.~\ref{eq:Hpert}) does not mix momenta during a
transfer process. Consequently, we may analytically solve the equations of
motion and explicitly present the solutions in terms of the Heisenberg operators
at $t = 0$, thus yielding separately the time evolution in the form of
2$\times$2 unitary matrix for each $\veck$,
\begin{equation}
   \begin{pmatrix}
      \, \cak(t) \, \\ \\ \cbk(t)
   \end{pmatrix}
   = \exp \left( -\frac{\im \epsk t}{\hbar} \right)
   \begin{pmatrix}
      \coslk     & -\im \sinlk \\
     -\im \sinlk & \coslk
   \end{pmatrix}
   \begin{pmatrix}
      \, \cak(0) \, \\ \\ \cbk(0)
   \end{pmatrix}
   .
\end{equation}
In particular, the operators the average of which provides the occupation
numbers of the first subsystem are given by
\begin{multline}
   \cakd(t) \, \cak(t) \, =
   \ccoslk \, \cakd(0) \, \cak(0) + \ssinlk \, \cbkd(0) \, \cbk(0) \\
   -\im \coslk \sinlk
   \left[ \cakd(0) \, \cbk(0) - \cbkd(0) \, \cak(0) \right]
   \label{eq:n1kt} 
\end{multline}
whereas a similar expression follows for $\cbkd(t) \, \cbk(t)$.
As all time dependence is in the operators, the Heisenberg picture only requires
the explicit knowledge of the initial density matrix $\rhoh_0$ to calculate any
expectation values at arbitrary later times. In general, a time dependent
observable $Y_t$ representing the time dependent ensemble average of a
Heisenberg operator $\hat{Y}(t)$ is to be extracted according to the recipe
\cite{kreuzer1981}
\begin{equation}
   Y_t = \fbraket{\hat{Y}(t)} = \Sp \left[ \rhoh_0 \hat{Y}(t) \right].
\end{equation}
As both fermion reservoirs are assumed to be uncoupled at $t = 0$ and before,
the initial density matrix $\rhoh_0$ reduces to the product of the two reservoir
density matrices
\begin{align}
   \rhoh_0 
   & = \frac{1}{Z_0}
       \exp \left( -\beta \left( \Hunperta - \muup_1 \Nop_1 \right) \right) \,
       \exp \left( -\beta \left( \Hunpertb - \muup_2 \Nop_2 \right) \right)
       \nonumber \\
   & = \frac{1}{Z_0}
       \prod_{\veck}
       \exp \left( -\beta \left( \epsk - \muup_1 \right) \cakd \cak \right) \,
       \exp \left( -\beta \left( \epsk - \muup_2 \right) \cbkd \cbk \right),
   \label{eq:rho0}
\end{align}
where $\beta = 1 / \kB T$, $Z_0$ is the partition function and
$\muup_1, \muup_2$ denote the reservoir chemical potentials compatible with
the initial densities $n_{10}$ and $n_{20}$. The number operators counting the
fermions in the distinghuised reservoirs are given by
\begin{equation}
   \Nop_{\! j} = \sumk \cjkd \cjk, \quad j = 1, 2.
\end{equation}
Obviously, the initial density matrix (\ref{eq:rho0}) translates to the
Fermi-Dirac distribution ruling the initial occupation numbers:
\begin{equation}
   \fbraket{\clkd(0) \, \cjk(0)} = \Fj(\epsk) \, \delta_{l, j}, \quad
   \Fj(E) \equiv
   \frac{1}{1 + \exp \, \left( \beta (E - \muup_j) \right)}, \quad j, l = 1, 2.
   \label{eq:cc0}
\end{equation}
Combining Eqs.~(\ref{eq:n1kt}) and (\ref{eq:cc0}), we arrive at
\begin{equation}
   \fbraket{\cakd(t) \, \cak(t)} \, =
   \ccoslk \, F_1(\epsk) + \ssinlk \, F_2(\epsk), \label{eq:n1kav} \\
\end{equation}
In particular, we obtain the generic expression for the time dependent fermion
density in the first reservoir from
\begin{align}
   n_1(t)
   & = \frac{1}{L_x L_y} \fbraket{\Nop_1(t)} =
       \frac{1}{L_x L_y} \sumk \fbraket{\cakd(t) \, \cak(t)} \nonumber \\
   & = \frac{1}{L_x L_y} \sumk
       \left[ \ccoslk \, F_1(\epsk) + \ssinlk \, F_2(\epsk) \right],
\end{align}
which can be rewritten as
\begin{equation}
   n_1(t) = \half \left( n_{10} + n_{20} \right) +
            \frac{1}{2 L_x L_y} \sumk \cosllk \,
            \left[ F_1(\epsk) - F_2(\epsk) \right]
   \label{eq:n1t}
\end{equation}

\section{Solution and discussion}
\label{sec:soldis}
Defining the diffusion current density $\JD(t)$ to be the rate of change of
the fermion density in the first reservoir, we may obtain the corresponding
expression for the present model by simply taking the time derivative of
Eq.~(\ref{eq:n1t}),
\begin{equation}
   \JD(t) = \frac{\d n_1(t)}{\d t}
          = -\frac{1}{\hbar L_x L_y} \sumk \lkv \sinllk \,
            \left[ F_1(\epsk) - F_2(\epsk) \right]
   \label{eq:JDt}
\end{equation}
It is worth noting that the entropy production is unambigously determined by
the total diffusion current
$\ID = \JD L_x L_y = \d \fbraket{\Nop_1(t)} / \, \d t$,
the initial difference of the chemical potentials $\muup_1 - \muup_2$ and the
common, ambient temperature $T$. Starting from the time dependent density
matrix $\rhoh_t$ satisfying the quantum Liouville equation
\begin{equation}
    \im \hbar \, \frac{\d \rhoh_t}{\d t} = \left[ \Htot, \rhoh_t \right],
    \label{eq:liouville}
\end{equation}
we first consider the relative entropy
\cite{ojima1987,ichiyanagi1989-1,ichiyanagi1989-2,ichiyanagi1989-3,kawai2007}
\begin{equation}
   S[\rhoh_t, \rhoh_0] =
   -\kB \Sp \left[ \rhoh_t \left( \ln \rhoh_0 - \ln \rhoh_t \right) \right]. 
   \label{eq:relent}
\end{equation}
From Eq.~(\ref{eq:liouville}) one may easily derive the well-known result 
$\Sp (\rhoh_t \ln \rhoh_t) = \Sp (\rhoh_0 \ln \rhoh_0)$ which is the
mathematical statement of the Shannon entropy failing to evolve in time.
Substitution into Eq.~(\ref{eq:relent}) yields
\begin{equation}
   S[\rhoh_t, \rhoh_0] =
   -\kB \Sp \left[ \left( \rhoh_t - \rhoh_0 \right) \ln \rhoh_0 \right], 
\end{equation}
whereas the factorized, initial density matrix in Eq.~(\ref{eq:rho0}) leads to
\begin{equation}
   \Sp \left( \rhoh_t \ln \rhoh_0 \right)
   = -\beta
   \Sp \left[
             \rhoh_t \left( \Hunpert - \muup_1 \Nop_1 - \muup_1 \Nop_2 \right)
       \right].
   \label{eq:relent1}
\end{equation}
Next, we use
$\rhoh_t = \exp \, (-\im \Htot t / \hbar) \, \rhoh_0
\exp \, (\im \Htot t / \hbar)$ which provides the formal solution to the
Liouville equation, to eliminate the time dependence of the density matrix
and recast Eq.~(\ref{eq:relent1}) in the Heisenberg picture:
\begin{equation}
   \Sp \left( \rhoh_t \ln \rhoh_0 \right)
   = -\beta \,
   \Sp \left[
             \rhoh_0
             \left( \Hunpert(t) - \muup_1 \Nop_1(t) - \muup_1 \Nop_2(t) \right)
       \right].
   \label{eq:relent2}
\end{equation}
Using Eq.~(\ref{eq:relent2}) and bearing in mind that $\Hunpert$ is a constant
of motion because the diffusion process is not accompanied by energy exchange,
we obtain
\begin{equation}
   S[\rhoh_t, \rhoh_0] = -\frac{1}{T} \sum_{j = 1, 2}
   \left[
         \muup_j \left( \fbraket{\Nop_j(t)} - \fbraket{\Nop_j(0)} \right)
   \right].
   \label{eq:relent3}
\end{equation}
Eq.~(\ref{eq:relent3}) reveals that any entropy change is to be considered
configurational in the sense that it is exclusively due to the
redistribution of all fermions over both reservoir areas.
Emerging as the time derivative of the relative entropy, the entropy production
can be directly derived from Eq.~(\ref{eq:relent3}):
\begin{equation}
   \frac{\d S(t)}{\d t} = -\frac{1}{T} \sum_{j = 1, 2}
   \left[
         \muup_j \, \frac{\d \fbraket{\Nop_j(t)}}{\d t}
   \right].
\end{equation}
Finally, taking into account that the total number of fermions is a constant of
motion as well, we arrive at
\begin{equation}
   \frac{\d S(t)}{\d t} = -\frac{1}{T} \left( \muup_1 - \muup_2 \right) \ID(t).
\end{equation}
Hence, the entropy production is instantly proportional to the diffusion current
while the sign of the latter depends on which of either reservoirs is being
emptied into the other.
In order to obtain more detailed results or numerical output based on
Eq.~(\ref{eq:n1t}), it is paramount to specify the functional dependence of
$\lk$ on the wave vector $\veck$ as well as to identify the range of allowed
wave vectors. A simple and straighforward model favoring the transfer of
high-energy fermions corresponds to $\lk$ being chosen proportional to $\epsk$,
say
\begin{equation}
   \lkv = \lk = \lambdaup \epsk, \quad \lambdaup > 0
\end{equation}
and will be further explored as a benchmark.
Moreover, it follows from Eq.~(\ref{eq:n1t}) emerging as a sum of
trigonometric oscillations that the time evolution of the present diffusion
model is bound to be reversible as long as the allowed wave vectors run through
a finite, countable set. Phrased otherwise, time irreversibility is expected
to occur only if the thermodynamic limit is taken with
$L_x, L_y, \fbraket{\Nop_1}$, $\fbraket{\Nop_2} \to \infty$, the densities
$n_1$ and $n_2$ however remaining finite. Traditionally, this encompasses the
conversion of the summation over $\veck$ into a 2D integration according to
\begin{equation}
   \frac{1}{L_x L_y} \sumk \to \frac{1}{4 \pi^2} \int \d^2 k
\end{equation}
Carrying out the latter and transforming to the ``polar'' coordinates
$(\! \sqrt{E}, \, \phi)$ according to
\begin{equation}
   k_x = \frac{\sqrt{2 m E}}{\hbar} \cos \phi, \;
   k_y = \frac{\sqrt{2 m E}}{\hbar} \sin \phi, \quad
   E \geqslant 0, \; 0 \leqslant \phi \leqslant 2 \pi, 
\end{equation}
we obtain
\begin{align}
   n_1(t)
   & = \half \left( n_{10} + n_{20} \right) +
       \frac{1}{8 \pi^2} \! \int \! \d^2 k \,
       \cos \left( \frac{2 \lambdaup \epsk t}{\hbar} \right) \,
       \left[ F_1(\epsk) - F_2(\epsk) \right] \nonumber \\
   & = \half \left( n_{10} + n_{20} \right) +
       \frac{m}{8 \pi^2 \hbar^2} \! \int_0^{\infty} \!\!\!\! \d E \!
       \int_0^{2 \pi} \!\!\!\! \d \phi \,
       \cos \left( \frac{2 \lambdaup E t}{\hbar} \right) \,
       \left[ F_1(E) - F_2(E) \right] \nonumber \\
   & = \half \left( n_{10} + n_{20} \right) +
       \frac{m}{4 \pi \hbar^2} \! \int_0^{\infty} \!\!\!\! \d E \,
       \cos \left( \frac{2 \lambdaup E t}{\hbar} \right) \,
       \left[ F_1(E) - F_2(E) \right]. \label{eq:n1int}
\end{align}
Clearly, the damped oscillations are driven by the initial chemical potential
difference $\muup_1 - \muup_2$. Note that the 2D geometry of the two reservoirs
enables us to extract analytically  the relation between $\muup_1$, $\muup_2$
and the corresponding initial densities from
\begin{equation}
   \muup_j = \kB T \ln
   \left[ \exp \left( \frac{2 \pi \hbar^2 n_{j0}}{m \kB T} \right) - 1 \right],
   \qquad j = 1, 2.
\end{equation}
Any further analytical treatment of Eq.~(\ref{eq:n1int}) amounts to either
a series expansion of the Fermi-Dirac distribution function when it comes to
keep track of the temperature dependence, or to studying the extreme quantum
limit at zero temperature. In the latter case, the Fermi-Dirac functions reduce
to step functions respectively imposing integration boundaries at
$E = \muup_1 > 0$ and $E = \muup_2 > 0$, yielding
\begin{align}
   n_1(t)
   & = \half \left( n_{10} + n_{20} \right) +
       \frac{m}{4 \pi \hbar^2} \! \int_{\muup_2}^{\muup_1} \!\!\!\! \d E \,
       \cos \left( \frac{2 \lambdaup E t}{\hbar} \right) \nonumber \\
   & = \half \left( n_{10} + n_{20} \right) +
       \frac{m}{8 \pi \hbar \lambdaup} \cdot \frac{1}{t}
       \left[
             \sin \left( \frac{2 \lambdaup \muup_1 t}{\hbar} \right)
            -\sin \left( \frac{2 \lambdaup \muup_2 t}{\hbar} \right)
       \right]
\end{align}
On the other hand, in the classical limit corresponding to
$\kB T \gg \pi \hbar^2 n_{j0} / m$ and $\muup_1, \muup_2 < 0$ we may replace
the Fermi-Dirac functions with bare exponentials to arrive at
\begin{align}
   n_1(t)
   & = \half \left( n_{10} + n_{20} \right) + \frac{m}{4 \pi \hbar^2}
       \left(
             \e^{\, \beta \hskip 0.2 mm \muup_1}
            -\e^{\, \beta \hskip 0.2 mm \muup_2}
       \right)
       \!  \int_0^{\infty} \!\!\!\! \d E \,
       \cos \left( \frac{2 \lambdaup E t}{\hbar} \right) \, \e^{-\beta E}
       \\ \nonumber
   & = \half \left( n_{10} + n_{20} \right) + \frac{\beta m}{4 \pi}
       \left(
             \e^{\, \beta \hskip 0.2 mm \muup_1}
            -\e^{\, \beta \hskip 0.2 mm \muup_2}
       \right)
       \frac{1}{\beta^2 \hbar^2 + 4 \lambda^2 t^2} \\ \nonumber
   & = \half \left( n_{10} + n_{20} \right) +
       \half \left( n_{10} - n_{20} \right)
       \frac{\beta^2 \hbar^2}{\beta^2 \hbar^2 + 4 \lambda^2 t^2}
\end{align}
This time, the evolution of $n_1(t)$ towards its steady-state value
$n_{\mathrm S} = 1/2 (n_{10} + n_{20})$ is governed by a $t^{-2}$ power law,
while no oscillations are observed during the transient regime. 
Note that, whenever irreversibility is established, the evolution from the
original equilibrium state to the final steady state corresponds to an entropy
increase because $n_{10}$, the original density in reservoir 1, decreases
(increases) towards $n_{\mathrm S}$ provided that $\muup_1 > ( < ) \, \muup_2$,
as can also be seen from the entropy change per particle,
\begin{equation}
   \Delta s \equiv \lim_{t \to \infty}
   \frac{S[\rhoh_t, \rhoh_0]}{\fbraket{\Nop_1(t)} + \fbraket{\Nop_2(t)}} =
   \frac{\muup_1 - \muup_2}{T} \cdot \frac{n_{10} - n_{20}}{n_{10} + n_{20}}.
\end{equation}

\section{Conclusion}
\label{sec:concl}
The quantum diffusion model presented in this paper can be solved analytically
to obtain a closed-form integral representation of the time dependent
observables, such as the fermion density in the reservoirs. The long-time
behavior is governed by a power law $t^{-\alpha}$ where $\alpha = 1$ and
$\alpha = 2$ respectively characterize the extreme quantum regime and the
classical regime. The time dependent entropy production is proportional to the
diffusion current while the irreversibility of the latter in the case of the
thermodynamic limit corresponds to a net increase of the (configurational)
entropy when the steady-state is attained.

\end{document}